\documentclass{aastex}
\usepackage{emulateapj5,times,mathptm}
\journalinfo{The Astrophysical Journal, 2002, astro-ph/0202080}

\newcommand{\gs}{{_>\atop^{\sim}}}

\slugcomment{ }

\shorttitle{X--RAY BRIGHT OPTICALLY QUIET GALAXIES}
\shortauthors{COMASTRI ET AL.}

\begin{document}


\title{The HELLAS2XMM survey: II. multiwavelength \\
 observations of P3: an X--ray bright, optically inactive 
galaxy\altaffilmark{1}}


\author{A. Comastri, M. Mignoli, P. Ciliegi}
\affil{Osservatorio Astronomico di Bologna,  
    via Ranzani 1, I--40127 Bologna, Italy}
\email{comastri,mignoli,ciliegi@bo.astro.it}
\author{P. Severgnini}
\affil{Dipartimento di Astronomia e Scienze dello Spazio, 
 Largo E. Fermi 5, I--50125, Firenze, Italy}
\email{paolas@arcetri.astro.it}
\author{R. Maiolino}
\affil{Osservatorio Astrofisico di Arcetri, 
 Largo E. Fermi 5, I--50125, Firenze, Italy}
\email{maiolino@arcetri.astro.it}
\author{M. Brusa}
\affil{Dipartimento di Astronomia Universita' di Bologna, 
    via Ranzani 1, I--40127 Bologna, Italy}
\email{brusa@bo.astro.it}
\author{F. Fiore}
\affil{Osservatorio Astronomico di Roma, 
    via Frascati 33, I--00040 Monteporzio, Italy}
\email{fiore@quasar.mporzio.astro.it}
\author{A. Baldi, S. Molendi}
\affil{Istituto di Fisica Cosmica -- CNR,  
 via Bassini 15, I--20121, Milano, Italy}
\email{baldi,silvano@ifctr.mi.cnr.it}
\author{R. Morganti}
\affil{Netherlands Foundation for Research in Astronomy, 
      Postbus 2, 7990 AA Dwingeloo, The Netherlands}
\email{morganti@nfra.nl}
\author{C. Vignali}
\affil{Department of Astronomy and Astrophysics, 
 The Pennsylvania State University, 525 Davey Lab, 
University Park, PA 16802, USA}
\email{chris@astro.psu.edu}
\author{F. La Franca, G. Matt, G.C. Perola}
\affil{Dipartimento di Fisica Universita' di Roma Tre,
    via della Vasca Navale 84, I--00146 Roma, Italy}
\email{matt,lafranca,perola@fis.uniroma3.it}

\altaffiltext{1}{This work is based on observations collected at the European 
Southern Observatory, La Silla and Paranal (proposal 66.B-0472(A)), 
Chile and observations 
made with the XMM--{\it Newton}, an ESA science mission with instruments 
and contributions directly funded by ESA member states and the USA (NASA).}




\begin{abstract}
Recent X--ray surveys have clearly demonstrated that a population 
of optically dull, X--ray  bright galaxies is emerging at 2--10 keV 
fluxes of the order of $10^{-14}$ erg cm$^{-2}$ s$^{-1}$.
Although they might constitute an important fraction 
of the sources responsible for the hard X--ray background,  
their nature is still unknown. With the aim to better understand 
the physical mechanisms responsible for the observed properties,  
we have started an extensive program of multiwavelength follow--up
observations of hard X--ray, optically quiet galaxies
discovered with XMM--{\it Newton}.  
Here we report the results of what can be considered the first 
example of this class of objects: CXOUJ031238.9--765134, originally 
discovered by {\it Chandra}, and   
optically identified by Fiore et al. (2000) with an apparently normal 
early--type galaxy at $z$=0.159, usually known as ``{\tt FIORE P3}''. 
The analysis of the broad-band energy distribution 
suggests the presence of a heavily obscured active nucleus.


\end{abstract}


\keywords{galaxies: active --- galaxies: individual (P3) --- galaxies: nuclei
---  X--rays: galaxies}


\section{Introduction}

The discrete sources responsible for a large fraction ($>$75 \%) 
of the X--ray background emission below 8 keV have been detected 
by {\it Chandra} deep surveys down to extremely faint fluxes   
S$_{2-8 keV} <$ 10$^{-15}$ erg cm$^{-2}$ s$^{-1}$ 
(Mushotzky et al. 2000; Giacconi et al. 2001; Hornschmeier et al. 2001; 
Brandt et al. 2001; Tozzi et al. 2001).
The unprecedented arcsec {\it Chandra} spatial resolution 
allows to unambiguously identify the optical counterparts
of the X--ray sources.
Extensive programs of optical identifications  
showed that about half of them are associated with optically bright
objects at redshifts $<$ 1.5 while the other half appears to be a mixture
of optically faint (I $>$ 23.5) galaxies and  Active Galactic Nuclei,  
presumably at high redshift. 
The source breakdown of the optically bright population 
is intriguing. Contrary to the situation for the faint soft X--ray selected
ROSAT sources (Schmidt et al. 1998), only a minority ($<$ 20\%)
of them are associated with broad line objects and, even more surprisingly,   
some 50\% of the spectroscopically identified sources do not show  
obvious signatures of AGN activity in their optical spectra, which are 
instead typical of early-type ``normal'' galaxies (Hornschemeier et al. 2001;
 Giacconi et al. 2001; Barger et al. 2001).
These findings seem to support previous claims (Griffiths et al. 1995) 
based on lower spatial resolution ROSAT observations. 
  
The large X--ray--to--optical flux ratio, which exceeds by more than 
one order of magnitude the average value of early--type galaxies 
of similar optical luminosity (Fabbiano, Kim, \& Trinchieri 1992), 
and the hard X--ray spectra, determined from the analysis of X--ray colors,  
both suggest that (obscured) AGN activity is taking place in their nuclei.
In principle X--ray spectroscopy could provide 
a stringent test on their nature.  
Unfortunately, all the sources are detected 
with a number of photons which is too small
to apply conventional X--ray spectral fitting techniques and 
to constrain the absorbing column density.
The lack of optical emission lines could be also explained 
if the nuclear light is overshined by either the stellar continuum or 
a non--thermal component, or if they are not efficiently produced. 

A much better understanding of the sources powering X--ray bright, 
optically quiet galaxies can be achieved only by means of multiwavelength
observations. 
For this reason we have started an extensive program of 
multifrequency follow--up observations of hard X--ray selected sources 
serendipitously discovered in XMM--{\it Newton} fields
over an area of about 3 deg$^2$ (the XMM High Energy Large Area Survey:
{\tt HELLAS2XMM}; see Baldi et al. 2002). 

In this paper we present a multiwavelength study of what can be considered
the first example of these objects:  
the X--ray source CXOUJ 031238.9--765134,  
optically identified as a bright ($R$=18.0), 
``normal'' early-type galaxy at $z$=0.159
by Fiore et al. (2000, hereinafter F00) in their pilot study of 
serendipitous X--ray sources in two shallow {\it Chandra} fields.

\section{Observations}

The field surrounding the radio loud quasar PKS~0312--76 
has been observed by both {\it Chandra} and XMM--{\it Newton}
during their Calibration and Performance Phases. 
The {\it Chandra} source CXOUJ 031238.9--765134
(hereinafter P3, being the third source catalogued by F00
in the PKS field) was also detected by XMM with a number of 
photons sufficient for a rough spectral analysis. 
Here we present the X--ray spectral analysis combined 
with deep radio observations and near--infrared and optical spectroscopy.

\subsection{X--ray data} 

The analysis of {\it Chandra} data has been already reported 
by F00 based on preliminary calibration data 
and detection techniques. An additional observation of the 
PKS~0312--76 field was retrieved from the archive, combined with the
first one and analyzed using version 2.2 of the CXC software.
The periods of high background were filtered out leaving about 24.7 ksec
of useful data. 
The {\tt WAVDETECT} algorithm (Freeman et al. 2002)
was run on the cleaned image conservatively setting 
the false-positive probability threshold to 10$^{-7}$.
The source was clearly detected in both the soft 
(0.5--2 keV) and hard (2--8 keV) bands with 40$\pm$6 and 22$\pm$5 
net counts, respectively. Although the counting statistics 
do not allow a proper spectral analysis, the hard to soft band ratio 
of 0.55$\pm$0.15 implies a hard power-law spectrum
($\Gamma \simeq$ 1.4) assuming the Galactic absorption column density 
($N_H$ = 8 $\times$ 10$^{20}$ cm$^{-2}$; Dickey \& Lockman 1990).
The 2--10 keV {\it Chandra} flux is 2.5 $\times$ 10$^{-14}$ 
erg cm$^{-2}$ s$^{-1}$, consistent with that reported by F00.

Our target was clearly detected also in the 25 ksec XMM--{\it Newton} 
PV observation of the same field.
Even though the XMM positional accuracy is not as good as the 
{\it Chandra} one, the two X--ray centroids agree within about 2 arcsec.
The XMM and {\it Chandra} contours are overlaid on a R-band optical image
in Fig.~1. 
\figurenum{1}
\centerline{\includegraphics[angle=0,width=9.5cm]{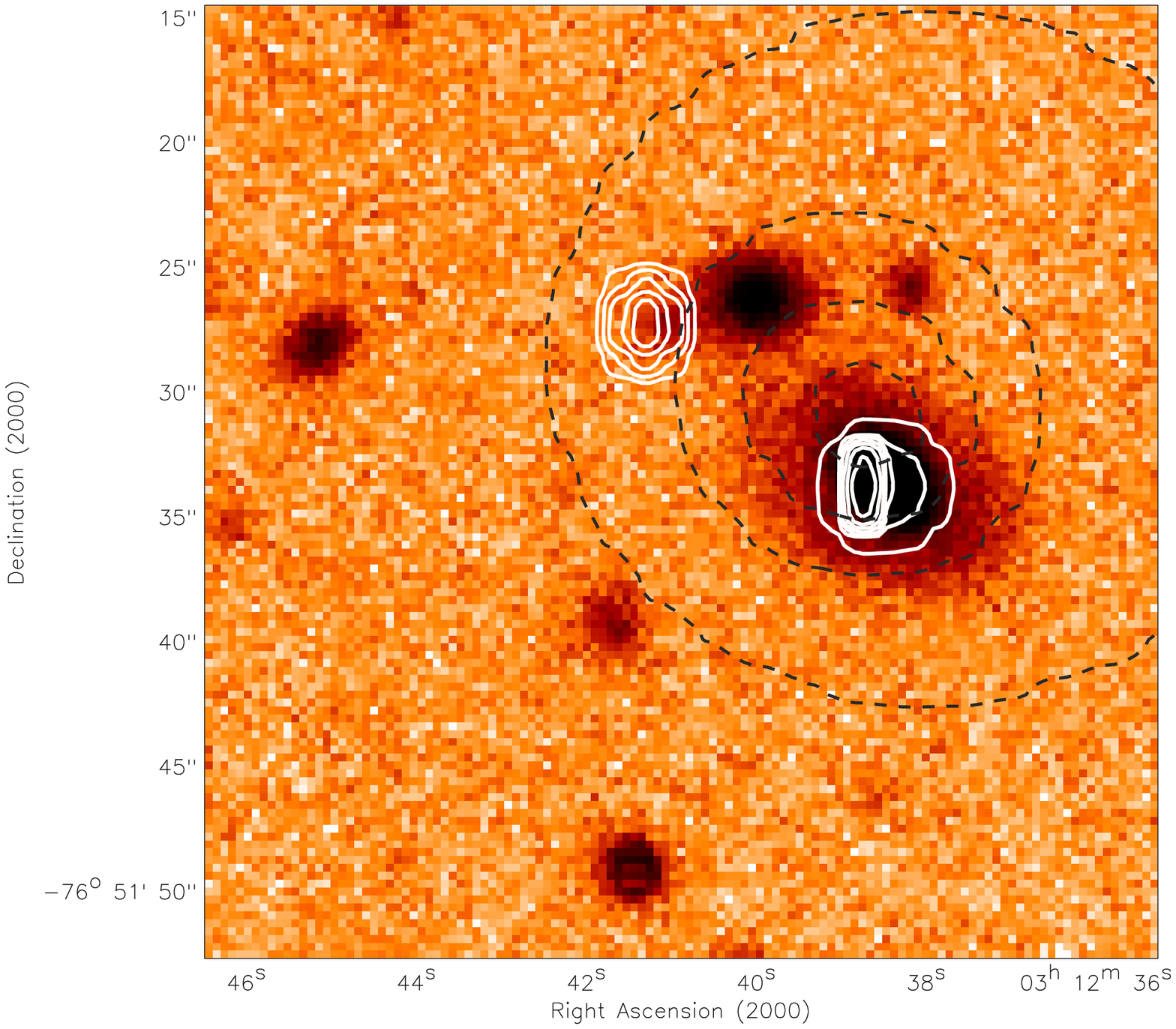}}
\figcaption{\footnotesize 
The {\it Chandra} (solid line) and XMM--{\it Newton} 
(dashed line) contours overlaid on the R band image. 
P3 is the galaxy coincident with the 
southern {\it Chandra} source. 
\label{fig1}}
\centerline{}
\centerline{}
\noindent 
It can be seen that the elongated XMM contours are probably 
due to the presence of a faint X--ray source which is clearly 
resolved by {\it Chandra} at a distance of $\sim$ 6 arcsec from P3.
This object appears to be coincident with an extremely faint 
R $>$ 24 optical counterpart. 
Its X--ray flux is only about 10--15 \% of P3, while its hardness 
ratio suggests a spectral shape similar to P3. 

The XMM data were processed using version 5.0 of the Science Analysis System 
(SAS). The event files were cleaned up from hot pixels and soft proton
flares (see Baldi et al. 2002 for details). The hot pixels were removed       
using both the XMM--SAS and the IRAF task {\it cosmicrays}, while for 
the proton flares a count rate threshold was applied removing 
all the time intervals with a count rate greater than 0.15 c/s in the 
10--12.4 keV energy range for the two MOS and greater than 0.35 c/s 
in the 10--13 keV band for the {\it pn} unit.
The resulting exposure times are 24.7, 26.5 and  26.1 ksec in the 
{\it pn}, MOS1 and MOS2 detectors, respectively.
The source is clearly detected in both the soft (0.5--2 keV)
and hard (2--10 keV) bands with an approximately equal 
number of net counts (about 100 for each band    
combining the PN and MOS data).
Source and background spectra were extracted from the cleaned events using an 
extraction radius of about 26'' for the {\it pn}, 35'' for MOS1 and
17'' for MOS2 (the source is nearby a CCD gap). Background spectra
were extracted by nearby source--free regions with radii between 1 and 1.5 
arcminutes depending on the detector.
The PN and MOS spectra were rebinned with at least 20 counts per channel 
and fitted simultaneously using {\tt XSPEC 11.0} 
leaving the relative normalizations
free to vary. The latest version of response and effective area files
was used. The Galactic column density in the direction of P3 
is included in the spectral fits. The limited counting statistics of 
the observed spectra allow us to fit only simple models.
Acceptable fits are obtained with both a power law
with photon index $\Gamma$ = 1.10$\pm$0.35 and
with a thermal model with an extremely high temperature  
$kT >$ 9 keV, confirming the presence of the hard spectrum inferred from 
the {\it Chandra} hardness ratio.
A good description of the data is also obtained fixing the 
power-law slope at $\Gamma$=1.8 (an average value for AGN; 
e.g., Nandra et al. 1997); this gives an absorption column density 
$N_H$ = 8$\pm$5 $\times$ 10$^{21}$ cm$^{-2}$ at the source frame
(errors are at the 90\% confidence level for one parameter).
The 2--10 keV flux is in the range 2.5--3.0 $\times$ 10$^{-14}$ 
erg s$^{-1}$ cm$^{-2}$, depending on the XMM detector, and is 
in good agreement with the {\it Chandra} value. 
The absorption-corrected 2--10 keV luminosity  
is about 3 $\times 10^{42}$ erg s$^{-1}$.
The good agreement between the power-law slope and the hard 
X--ray flux measured by {\it Chandra} and the XMM--{\it Newton} 
suggests the lack of significant flux and spectral variability 
on a time scale of about 7 months.
Taking into account the present uncertainties in the adopted spectral
models and the {\it Chandra} vs. XMM--{\it Newton} cross calibrations
(Snowden et al. 2002), the maximum fractional variability allowed by the data
is of the order of 30\%.
Our results are in good agreement 
with those recently obtained by Lumb, Guainazzi, \& Gondoin 
(2001) from the analysis of the same XMM--{\it Newton} data.
Our analysis, however, relies on a more detailed evaluation of the
instrumental and cosmic backgrounds at the source position.

\subsection{The Optical spectrum}

The spectroscopic observations have been performed with the ESO 3.6m
telescope equipped with {\tt EFOSC2} (Patat 1999) 
during two different observing runs.
On the night of 2000 January 4 we used the grism \#6 and a slit--width
of 2~arcsec, whereas the grism \#13 with a 1.5~arcsec slit was used
in the night of 2000 December 26. 
Both configurations \footnote {www.ls.eso.org/lasilla/Telescopes/360cat/efosc/docs/Efosc2Grisms.html} yield a similar
resolution of about 26 \AA\, but with the latter set--up the spectrum extends
more toward the red up to 9000 \AA\ .
The spectroscopic data have been reduced separately using standard
IRAF routines. Bias exposures taken on each night were stacked, checked for
consistency with the overscan regions of spectroscopic frames and subtracted
out. The bias--subtracted frames were then flat--fielded in a standard manner
using internal lamp flats obtained during the same run. The sky background
was removed by fitting a third--order polynomial along the spatial direction
in source free regions. The 2--dimensional spectra of P3 have been
checked by eye and extracted via optimal averaging (Horne 1986).
In both the observing runs the wavelength calibration was made using
the HeAr arc lamp, and using the
spectroscopic standard star LTT4364 (Hamuy et~al., 1994) for flux calibration.
The slit was oriented with different position angles during the two
observations: in one case we set the angle in order to include in the slit
both the principal target and a nearby fainter object (turned out to be
a background galaxy). The other position angle has been chosen to include a
bright star in the slit in order to perform an optimal subtraction of the
atmospheric telluric features. Due to the different slit orientations, the
non--photometric conditions of the nights and the relatively narrow slits we
used, the present spectra cannot be considered spectrophotometric. 
The comparison of spectra taken one year apart shows good
agreement, compatible with the precision of the flux calibration ($\sim$ 
15\%). 
In Fig.~2 we show the average of the two spectra, with the 
principal identified absorption lines labeled. We also mark the
[O{\tt II}] emission line at 3727 \AA\ and the expected position of the
H$\alpha$ features. Unfortunately, at the redshift of P3 the H$\alpha$ line
position coincides with the strongest atmospheric telluric band in the 
optical.
We did our best in order to remove the telluric features, but we
cannot absolutely exclude the presence of a weak feature, either in emission
or in absorption.
\figurenum{2}
\centerline{\includegraphics[angle=0,width=8.5cm]{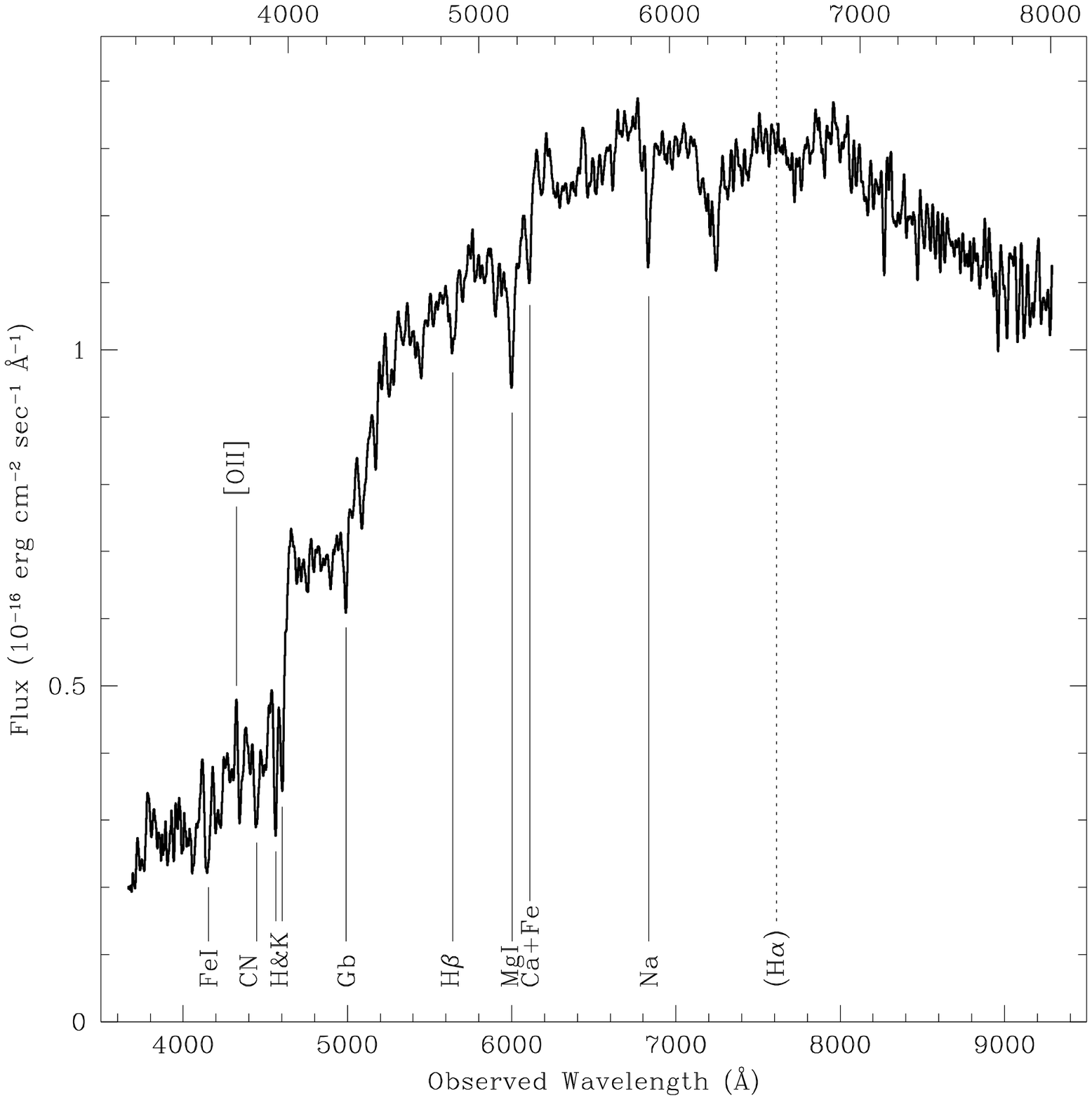}}
\figcaption{\footnotesize 
The flux calibrated co-added optical spectrum. The spectral features
are indicated with solid lines, while the dotted line
is at the wavelength corresponding to H$\alpha$. 
\label{fig2}}
\centerline{}
\centerline{}
The average redshift is $z$ = 0.1595$\pm$0.0007, 
measured from the position of the principal absorption lines and 
by cross-correlation with an early--type galaxy template.

\subsection{The near--infrared spectrum}

Near Infrared spectroscopic observations of the P3 galaxy were 
performed at the ESO Very Large Telescope (VLT).
The data were collected on 2000 November 10 and 11 under good 
seeing (0.7 to 0.9 arcsec) and photometric conditions. 
The galaxy has been observed with the 1\arcsec ~ slit in two
different filters, SH (1.42--1.83 $\mu$m) and SK (1.84--2.56 $\mu$m),
available for the Low Resolution grating (LR) in the Short Wavelength (SW)
configuration of {\tt ISAAC} (Moorwood et al. 1999).
The pixel size is 0.147 arcsec/pixel along the slit and 
the resolving powers were 500 and 450 in the SH and SK band, respectively.
The target was observed several times at different positions along the
slit (nodding mode). Each single exposure was 200~s long, resulting in
a total exposure time of 2 hours in each filter.\\
The data were reduced using routines from the {\tt ECLIPSE} (Devillard 1998), 
{\tt IRAF} and {\tt MIDAS} packages. Since we observed in nodding mode, 
the dark, bias, and sky contributions were removed by subtracting the 
pairs of offset frames.
The data were then flat-fielded using a master flat field image.
Wavelength calibration and correction for distortions along the slit were
performed using ArXe arc exposures, while the distortion along the spatial 
direction was corrected using the star-trace exposures provided by ESO.
The individual frames were then aligned and combined.
The telluric absorption features were removed using the spectroscopic 
standard Hip33590 (A0V) after deleting from the standard spectrum the 
intrinsic stellar features. To obtain the intrinsic shape of the continuum of
the galaxy and to perform the flux calibration we have 
used a blackbody curve with the appropriate temperature to fit
the intrinsic continuum of the standard.
In order to correct for the slit losses we have finally normalized the
spectra to the magnitudes in the H and Ks bands (H=15.7, K$^{\prime}$=14.9)
obtained by the acquisition images. 
The final spectra in both the SH and SK filters 
are shown in Fig.~3. The infrared spectrum is consistent,
within the precision of the flux calibration ($\sim$ 15\%),
with a featureless continuum. There is no evidence of significant 
emission  lines and in particular of P$\alpha$. 
\figurenum{3}
\centerline{\includegraphics[angle=0,width=8.5cm]{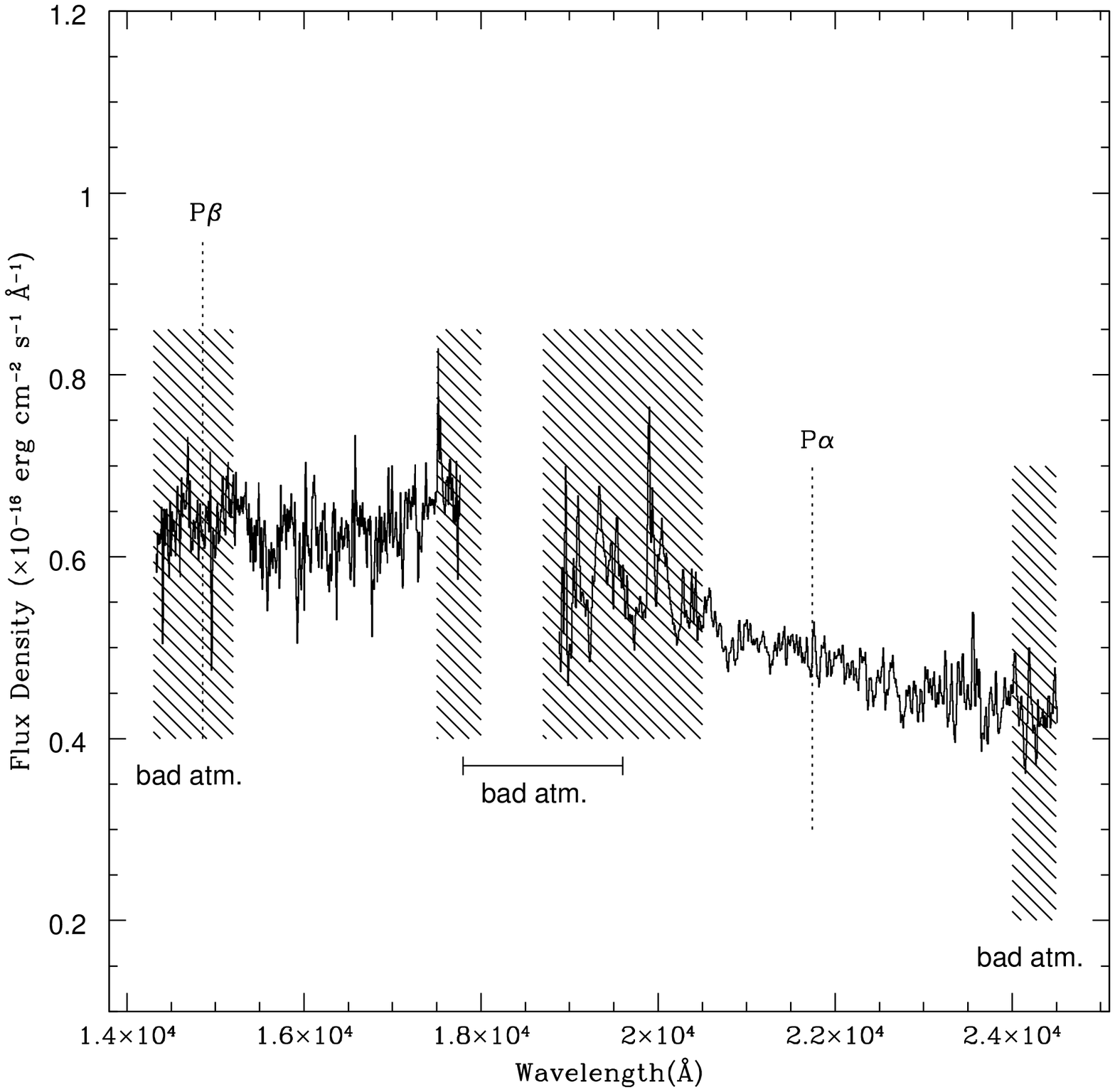}}
\figcaption{\footnotesize 
The VLT ISAAC infrared spectrum. The shaded regions correspond
to wavelengths of bad atmospheric transmissions. The expected 
positions of P$\alpha$ and P$\beta$ lines are also indicated. 
\label{fig3}}
\centerline{}
\centerline{}
\noindent
Imaging observations in the L band (3.8$\mu$m) have been 
performed at the VLT with  the ISAAC camera
along with the NIR infrared spectroscopy.
About half of the  total exposure time of 30 minutes 
was lost due to technical problems.
The images have been acquired combining chopping and telescope nodding
in opposite directions to allow a better sky subtraction (see ISAAC User
Manual).
All of the chopped images, i.e. background-subtracted frames with a
positive and negative image, have been flat fielded and then combined
together with a median filter to obtain the final image.
The data have been calibrated using the IR photometric standard HD219761
from Van der Bliek, Manfroid, \& Bouchet (1996) list.
The source is not detected at the limiting magnitude 
L = 13.4 (3$\sigma$ upper limit).

\subsection{Radio data}

Radio observations of the PKS~0312--76 field were performed at 5 GHz with
the Australia Telescope Compact Array (ATCA) in the 6-km configuration
(maximum baseline length), with a resolution of 2 $\times$ 2 arcsec$^2$.
The data were collected in a run of
12-hours on 2000 September 27. In order to improve the sensitivity
by a factor of $\sqrt{2}$, we used both ATCA receivers at 5000 MHz, each
with a bandpass of 128 MHz. The optimal positioning of the bands was obtained
by centering one frequency at 4800 MHz (4736-4864 MHz) and the other
at 5824 MHz (5760-5888 MHz). The primary flux density calibrator was
PKS~1934$-$638, while the source PKS~0230$-$790 was used as a phase and 
secondary amplitude calibrator. 
The field was observed in the mosaic mode by cycling through a grid 
of 5 pointings on the sky, one of which was centered
on the position of P3.  The grid of the pointings in the mosaic was
designed to yield an uniform noise over the area covered by the {\it Chandra}
data. With the mosaic technique, images obtained with single pointings are
combined together into a large image (mosaic) of the entire observed region.

The data were analyzed with the software package {\tt MIRIAD}. Each
bandpass (4800 and 5824 MHz) was calibrated and cleaned separately to
produce two individual images that we combined together into a single 
mosaic at the end of the reduction phase. Self--calibration was used 
to make additional correction to the antenna gains and to improve the 
image quality. The final map  has an uniform noise of
50 $\mu$Jy (1$\sigma$) over an area with a semicircular shape (due to
the odd numbers of pointings) with a radius of about 10 arcmin, surrounded
by an area where the noise increases with the distance from the
center. 

The P3 galaxy is well inside the area with an uniform noise of
50 $\mu$Jy. We searched for a possible radio source within a box
of 20 $\times$ 20 arcsec centered on the {\it Chandra} position.
No sources were found down to 100 $\mu$Jy (2$\sigma$).
Therefore we can conclude that the P3 galaxy has not a radio
counterpart in our map and that its 3$\sigma$ radio upper limit 
at 5312 MHz is 0.15 mJy.

\section{Discussion}

Although the overall spectral energy distribution 
(hereinafter SED) is dominated by the optical--infrared light
of the host galaxy, the X--ray flux level is 
almost two orders of magnitude higher 
than that expected on the basis of the $L_X$--$L_B$ correlation 
of early-type galaxies (Fabbiano et al. 1992). 

The relatively high X--ray luminosity and the X--ray spectral properties
strongly suggest nuclear activity in the central region of P3.
Based on the {\it Chandra} detection and the optical spectrum F00
suggested three different possibilities, namely: 
a radiatively inefficient advection dominated accretion flow (ADAF), 
a BL Lac object or a hidden, possibly obscured, Seyfert--like AGN. 
The multiwavelength observations presented in this paper allow us to 
investigate in more detail these possibilities.


The ASCA detection of relatively luminous 
($L_{2-10 keV} \simeq 10^{40}-10^{42}$ erg s$^{-1}$)
extremely hard power-law components ($\langle \alpha \rangle$ = 0.2) 
in the nuclei of six nearby elliptical galaxies which are known to 
host a massive black hole (Allen, Di Matteo, \& Fabian 2000) was consistent, 
given the observed radio emission, 
with an ADAF model in which a significant fraction of energy 
is removed from the flow by winds. 
This possibility however appears to be inconsistent 
with recent {\it Chandra} observations of three of these 
galaxies (Loewenstein et al. 2001): most of the spatially 
unresolved hard X--ray emission detected by ASCA originates 
in X--ray binaries.
The lack of significant hard X--ray emission at the position of the
optical nuclei allowed to place upper limits of about 1--2 
order of magnitude lower than the luminosities estimated by Allen et al. 
(2000). Although ADAF models with strong outflows or convection  
might still be accommodated with the {\it Chandra} observations, 
the parameters computed by Allen et al. (2000) to fit 
the ASCA data need to be reconsidered.
A detailed comparison of ADAF models to the present data is 
beyond the scope of this paper. Here we limit ourselves to compare  
the observed fluxes in the radio and X--ray band with 
the model spectra of Quataert \& Narayan (1999) for several values 
of the physical parameters involved in the ADAF models with winds
as most of the bolometric luminosity predicted by these models 
is produced  in the hard X--ray and radio bands.
For the range of model parameters computed by Quataert \& Narayan (1999) 
the X--ray flux density is about 1--2 order of  
magnitude greater than the radio flux density (see their Fig.~7).
The observed lower limit on the X--ray to radio flux density   
of P3 ($>$ 10$^{3}$) is clearly inconsistent with ADAF model predictions.
\begin{figure*}[t]
\figurenum{4}
\includegraphics[angle=0,width=8.5cm]{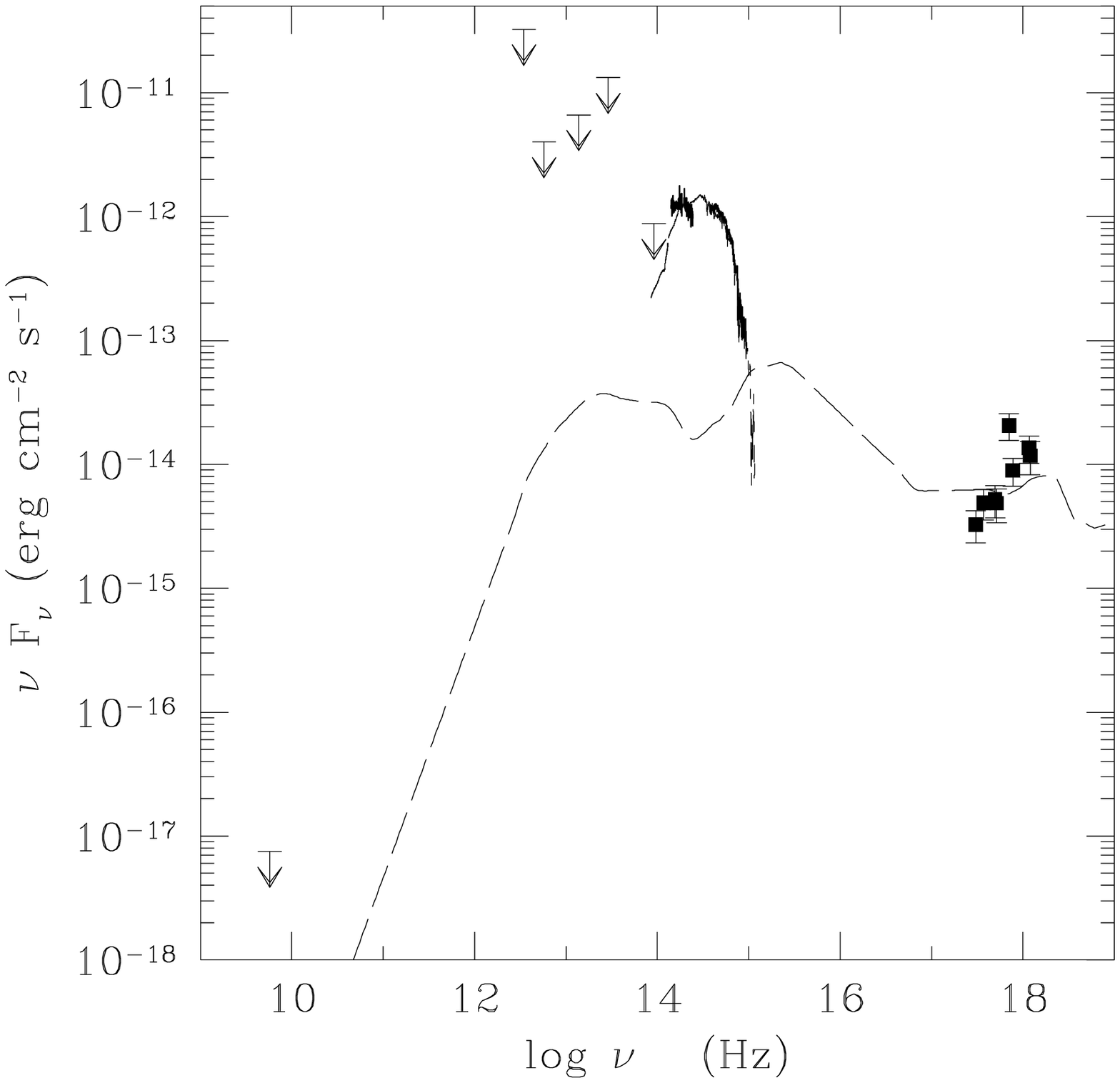}
\hspace{1.2cm} \ \
\includegraphics[angle=0,width=8.5cm]{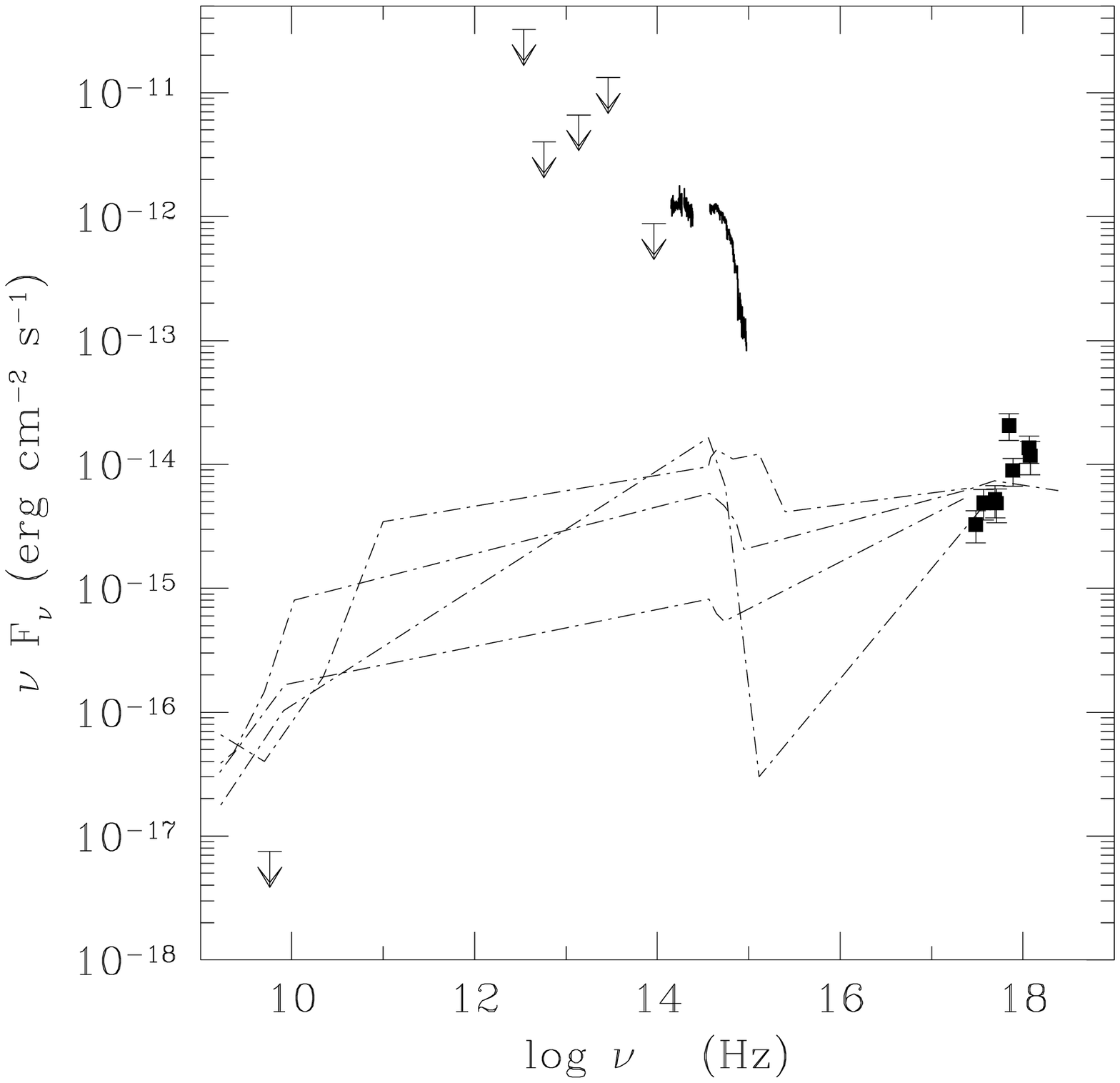}
\figcaption{\footnotesize 
Left panel: The broad-band spectral energy distribution of 
P3, including the upper limits in the 12$\mu$, 25$\mu$, 60$\mu$ and 
100$\mu$ IRAS bands (retrieved from the NED database), is compared 
with the average SED of a radio--quiet AGN (long dashed line) 
plus an early-type galaxy template (solid line). Right panel: 
the SED of the four low-luminosity AGN in the 
Ho (1999) sample (dot--dashed lines) 
hosted by an E--galaxy (NGC 6251, NGC 4261, M84 and M87) 
normalized to the P3 X--ray flux. 
\label{fig4}}
\end{figure*}
\noindent
The observed SED is consistent with an early-type galaxy template
normalized to match the optical and infrared spectrum plus the average 
SED of radio--quiet AGN (Elvis et al. 1994) rescaled to the observed
X--ray flux (Fig.~4, left panel). 
We note that the average SED of radio quiet quasars has been obtained 
from a sample of high-luminosity quasars and thus is unlikely to be 
appropriate for low-luminosity AGN (hereinafter LLAGN) 
which are known to be characterized
by a quite different broad-band spectrum (Ho 1999). In particular,  
low-luminosity objects lack an ultraviolet excess and have a more 
pronounced contribution from X--rays to the overall energy output.
The relative ratio between X--ray and optical flux is usually parameterized
by the two point spectral index $\alpha_{ox}$ 
(the slope obtained connecting the fluxes at 2500 \AA\ and 2 keV) 
which is significantly lower for low-luminosity AGN (between 0.5 and 1.0)
with respect to higher-luminosity Seyferts and quasars, for which
 $\langle \alpha_{ox} \rangle \simeq$ 1.5 (e.g., Brandt, Laor, \& Wills 2000).
The calculation of $\alpha_{ox}$ for P3 is not straightforward 
as the 2500 \AA\ flux density obtained extrapolating the observed optical
spectrum is likely to be dominated by the host galaxy. As a consequence, 
the derived value of $\alpha_{ox}$ = 0.92 should be considered an 
upper limit on the ``nuclear'' optical to X--ray flux ratio.
Prompted by the similarities between LLAGN and the P3 multiwavelength data, 
the broad-band SED of the four LLAGN hosted by 
an elliptical galaxy in the Ho (1999) sample have been compared 
with that of P3 and normalized to the observed X--ray flux (Fig.~4, right 
panel). Although the comparison sample is very small, we note that 
the average flux in the radio band of 
low-luminosity AGN greatly exceeds the observed upper limit.
The general trend of increasing X--ray variability for decreasing 
X--ray luminosities (Nandra et al. 1997)  
would also be inconsistent with the lack of strong X--ray variability.

The BL Lac hypothesis might be tenable 
despite the presence of a large Calcium break (F00) and 
the low radio flux density. 
Indeed, it has been proposed (Giommi \& Padovani 1994) that BL Lac objects
discovered in radio and X--ray surveys are the two extremes 
of a single population differing by the frequency of the 
high--energy cut--off in their energy distribution which is related 
to the maximum energy of the synchrotron emitting electrons.
Moreover, there is increasing evidence that the SED of BL Lac objects and
Flat Spectrum Radio Quasars (Blazars) 
can be unified in a spectral sequence determined by the total luminosity 
(Fossati et al. 1998; Ghisellini et al. 1998). 
The objects populating the low--luminosity extreme of the sequence 
peak at high frequency (in the X--ray band) and are called HBL.
The upper limit on the 5 GHz radio luminosity ($<$ 10$^{39}$ 
erg s$^{-1}$), on the radio to X--ray spectral index ($\alpha_{rx} < 0.6$,
$F_{\nu} \propto \nu^{-\alpha}$) and the flat slope in the X--ray band  
would be consistent with the presence of a rather extreme member of the 
HBL class hosted by the P3 galaxy.
The lack of X--ray variability, however, may constitute a problem with this 
interpretation.

An alternative explanation would be the presence of a hidden 
AGN. There are several examples of (obscured) AGN discovered only 
by means of X--ray observations and not recognized as such 
by optical spectroscopy (see Matt 2001 and references therein). 
In order to test whether this is the case for P3, we have 
computed the expected optical magnitude in the R band assuming 
an average value of the optical to X--ray flux ratio 
typical of hard X--ray selected unobscured quasars which turned out to
be R=19.4. For a ``standard'' Galactic extinction curve 
(Savage \& Mathis 1979) the best-fit $N_H$ value obtained 
from the analysis described in Sect.~2.1 
corresponds to $A_R$ = 2.6. Extensive simulations have been carried out,
using the {\tt IRAF} program {\tt artdata},
to estimate if a nuclear  point--like source with R=22 
would have been detected in the P3 nucleus. 
The results indicate that a nuclear source fainter than 
$R\approx$21.5 could not be revealed with the quality of the 
present data which are limited by poor seeing conditions 
(about 1.8 arcsec). Although the present findings indicate that 
a mildly obscured AGN could be hosted by P3, we note that 
hard X--ray selected objects with $L_X > 10^{42}$ erg s$^{-1}$
are generally characterized 
(with some exceptions, e.g. Brandt et al. 1997) 
by a dust to gas ratio much lower 
than the Galactic standard (Maiolino et al. 2001). 
Taking these results at the face value the putative X--ray mildly obscured 
nucleus would have probably been detected in the optical band, 
though this conclusion is subject to some uncertainties on
the relation between X--ray and optical absorption.

\figurenum{5}
\centerline{\includegraphics[angle=0,width=8.5cm]{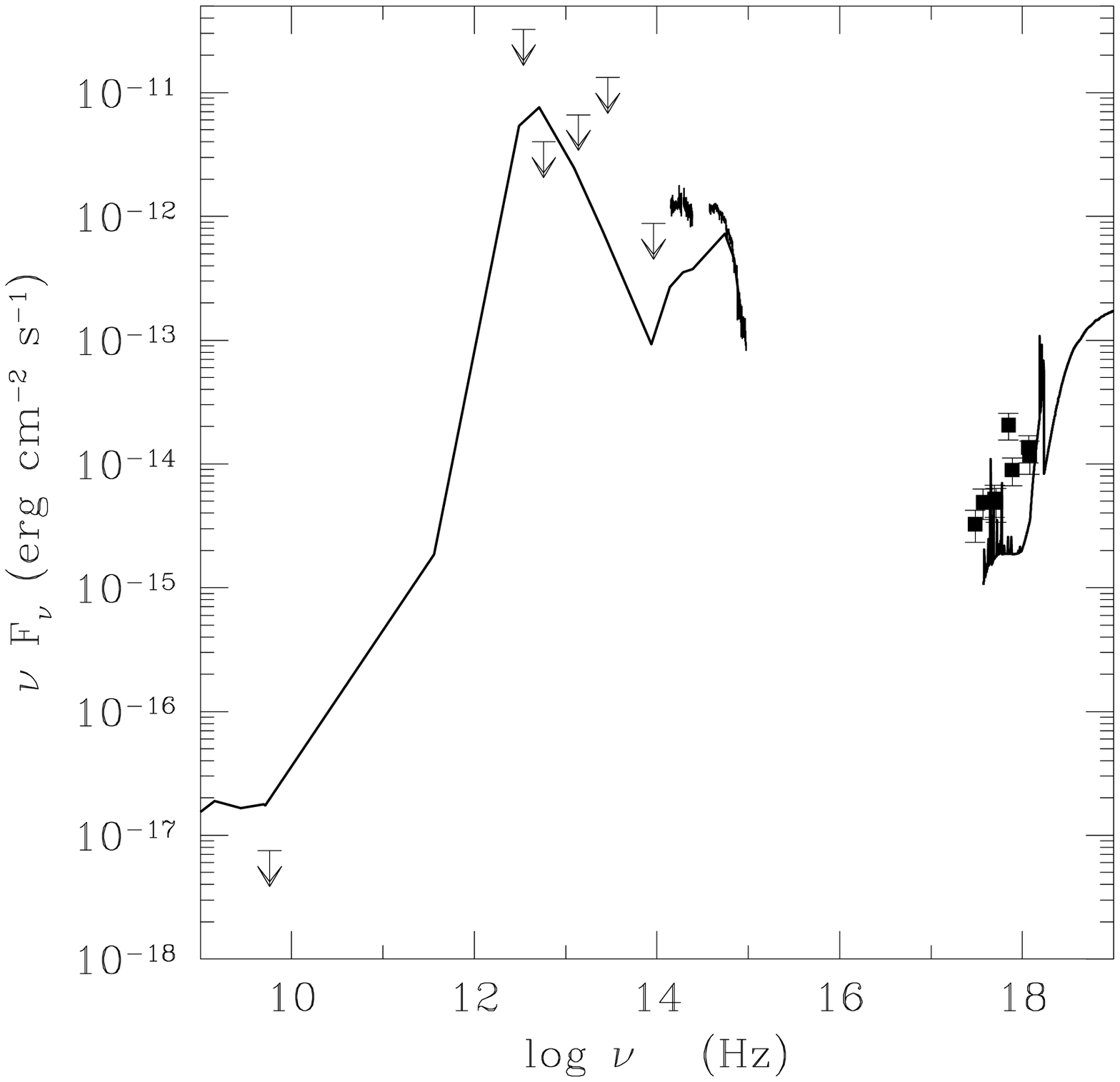}}
\figcaption{\footnotesize 
The observed SED is compared with that of the highly 
obscured Seyfert 2 galaxy NGC 6240. 
\label{fig5}}
\centerline{}
\centerline{}
\noindent
As a final possibility we have considered the hypothesis of 
a Compton thick ($N_H \gs$ 1.5 $\times$ 10$^{24}$ cm$^{-2}$) AGN. 
The P3 SED is compared (Fig.~5) with that 
of NGC 6240, a highly obscured AGN considered to be the prototype of 
this class of objects (Vignati et al. 1999).
The NGC 6240 SED is normalized to match the P3 optical flux.
Besides the disagreement in the radio band and at 60 $\mu$, which can be due 
to a more intense star--formation in NGC 6240, the P3 multiwavelength data 
are in relatively good agreement with the obscured AGN template.
If P3 hosts a Compton thick AGN with a SED similar to that of NGC 6240
then a luminous ($L_X \simeq 10^{44}$ erg s$^{-1}$) 
X--ray source should be present at higher energies.
In this case the observed X--ray emission would be due to a scattered 
nuclear component known to be common among highly obscured
AGN (Vignali et al. 2001 and references therein).
The lack of strong X--ray variability would support this hypothesis.

Whatever is the nature of the AGN powering P3, 
it is interesting to see how a source with similar broad-band 
properties would appear at high redshift and whether 
P3--like objects constitute a significant fraction of the 
X--ray faint sources which are being discovered in deep surveys
(Alexander et al. 2001). The 2--10 keV P3 luminosity can be detected, 
at the limit of the present {\it Chandra} surveys,
up to z$\simeq$1.5. At this redshift the optical and infrared
magnitudes of a passively evolving elliptical galaxy (Pozzetti et al. 1998) 
would be R $\simeq$ 24.2 and K$^{\prime} \simeq$ 18.5.
Such an object would not belong to the optically faint 
X--ray source population investigated by Alexander et al. (2001),
but rather to the class of Extremely Red Objects (EROs; R--K $>$ 5).
This is not surprising since passively evolving elliptical galaxies
have extremely red colors at $z > $1.
The EROs X--ray properties have been recently investigated 
using the 1 Ms exposure in the {\it Chandra} Deep Field North
(Alexander et al. 2002). The objects detected in the hard (2--8 keV) 
band have extremely flat spectral slopes and X--ray properties 
consistent with those expected from obscured AGN. 
It is also worth to remark that a high--z P3--like object would be bright 
enough in the optical band to allow spectroscopic observations at 8m 
class telescopes. Optical and infrared observations of faint X--ray 
sources might thus uncover several examples of completely hidden 
high--redshifts AGN.

\section{Conclusions}

The most important results obtained from our analysis of 
multiwavelength data can be summarized as follows:

$\bullet$ The high X--ray luminosity and hard X--ray spectrum 
clearly indicate the presence of nuclear activity in P3.

$\bullet$  There is no evidence of AGN emission lines both in the 
optical and infrared spectra. 

$\bullet$ The upper limit on the 5 GHz radio emission  
qualifies the AGN hosted by P3 (regardless of its origin) as a 
radio--quiet object.

$\bullet$ The multiwavelength properties allows us to rule out ADAF 
emission and the presence of a low-luminosity AGN with a SED similar 
to those of the few nearby objects in the Ho (1999) sample.

$\bullet$ An extreme BL Lac object, where the 
level of the non--thermal continuum in the radio and optical 
bands is very low, could provide an acceptable explanation 
of the P3 SED. The lack of X--ray variability, however, seems
to be at odd with this interpretation.

$\bullet$ The AGN responsible for the high X--ray luminosity
is likely to be hidden by a substantial 
(possibly Compton thick) column of cold absorbing gas.
If this is the case most of the observed X--ray emission would be due to 
a scattered/reprocessed nuclear component.

Sensitive broad band observations 
of a sample of X--ray bright, optically quiet galaxies 
are required to better understand their nature.
Deep optical and infrared spectroscopy of X--ray faint sources 
would allow to assess their fraction in deep X--ray surveys with 
important consequences for our understanding of the nature 
and the evolution of the X--ray background constituents.

\acknowledgments

We thank the entire {\it Chandra} team and in particular the CXC team 
for the support received in the data analysis.
This research has made use of the NASA/IPAC Extragalactic Database (NED)
which is operated by the Jet Propulsion Laboratory, California Institute 
of Technology, under contract with the National Aeronautics and Space 
Administration.  This paper used observations collected at the 
Australian Telescope Compact Array (ATCA), which is founded by the 
Commonwealth of Australia for operation as a National Facility by CSIRO.
The authors acknowledge partial support by the ASI contracts 
I/R/103/00 and I/R/107/00 and the MURST grant Cofin--00--02--36. 
CV also thanks the NASA LTSA grant NAG5--8107 for financial support.
We also thanks the referee for a fast and detailed report 
that improved the presentation of the results.

\clearpage






\begin{thebibliography}{}

\bibitem[]{} Alexander, D.M., Brandt, W.N., Hornschemeier, A.E., Garmire, 
G.P., Schneider, D.P., Bauer, F.E., \& Griffiths R.E. 2001, \aj, 122, 2156

\bibitem[]{} Alexander, D.M., Vignali, C., Bauer, F.E., Brandt, W.N., 
Hornschemeier, A.E., Garmire, G.P., \& Schneider, D.P. 2002, \aj, in press, 
(astro--ph/0111397)

\bibitem[]{} Allen, S.W., Di Matteo, T., \& Fabian, A.C. 2000, \mnras, 311, 493 
\bibitem[]{} Baldi, A., Molendi, S., Comastri, A., Fiore, F., Matt, G., 
\& Vignali, C.  2002, \apj, 564, 190

\bibitem []{} Barger, A.J., Cowie, L.L., Bautz, M.W., Brandt, W.N., Garmire, G.P., 
Hornschemeier, A.E., Ivison, R.J., \& Owen, F.N. 2001, \aj, 122, 2177

\bibitem[]{} Brandt, W.N., Fabian, A.C.,, Takahashi, K., Fujimoto, R., 
Yamashita, A., Inoue, H., \& Ogasaka, Y. 1997, \mnras, 290, 617 

\bibitem[]{} Brandt, W.N., Laor, A., \& Wills, B.J. 2000, \apj, 528, 637 

\bibitem[]{} Brandt, W.N., Alexander D.M., Hornschemeier A.E., et al. 2001, \aj, 122, 2810 
%
%
%
\bibitem[1998]{dev} Devillard, N. 1998, The Messenger, 87, 19
%
%
\bibitem[]{} Dickey, J.M., \& Lockman, F.J. 1990 \araa, 28, 215

\bibitem[]{} Elvis, M., et al. 1994, \apjs, 95, 1

\bibitem[]{} Fabbiano, G., Kim, D.W., \& Trinchieri, G. 1992, \apjs, 80, 531

\bibitem[]{} Fiore, F., et al. 2000, New Astronomy 5, 143 (F00)
%
%
\bibitem[]{} Fossati, G., Maraschi, L., Celotti, A., Comastri, A., \& Ghisellini, G. 
1998, \mnras, 299, 433  

\bibitem[]{} Freeman, P.E., Kashyap, V., Rosner, R., \& Lamb, D. 2002, \apjs, 138, 185

\bibitem[]{} Ghisellini, G., Celotti, A., Fossati, G., Maraschi, L., \& Comastri, A. 
1998, \mnras, 301, 451  

\bibitem[]{} Giacconi, R., et al. 2001, \apj, 551, 664  

\bibitem[]{} Giommi, P., \& Padovani, P. 1994, \mnras, 268, L51 

\bibitem[]{} Griffiths, R.E., Georgantopoulos, I., Boyle, B.J., Stewart, G.C., 
Shanks, T., \& Della Ceca, R. 1995, \mnras, 275, 77 

\bibitem[]{} Hamuy, M., et al. 1994, PASP, 106, 566

\bibitem[]{} Ho, L.C. 1999, \apj, 516, 672

\bibitem[]{} Horne, K. 1986, PASP, 98, 609
%
%
\bibitem[]{} Hornschemeier, A., et al. 2001, \apj, 554, 742 

\bibitem[]{} Loewenstein, M., Mushotzky, R.F., Angelini, L., Arnaud, K.A., 
\& Quataert, E. 2001, \apj, 555, L21 

\bibitem[]{} Lumb, D.H., Guainazzi, M., \& Gondoin, P. 2001, \aap, 376, 387  

\bibitem[]{} Maiolino, R., Marconi, A., Salvati, M., Risaliti, G., Severgnini, P., 
Oliva, E., La Franca, F., \& Vanzi, L. 2001, \aap,  365, 28

\bibitem[]{} Matt, G. 2001, in ``Issues in unification of AGNs'', ed. A. Marconi, 
R. Maiolino, \& N. Nagar (astro--ph/0107584)

\bibitem[]{} Moorwood, A.F.M., et al. 1999, The Messenger, 95, 1 
(www.hq.eso.org/instruments/isaac)

\bibitem[]{} Mushotzky, R.F., Cowie, L.L., Barger, A.J., \& Arnaud, K.A. 2000, Nature, 404, 459 

\bibitem[]{} Nandra, K., George, I.M., Mushotzky, R.F., Turner, T.J., 
\& Yaqoob, T. 1997, \apj, 476, 70 
%
%
\bibitem[]{} Patat, F. 1999 Efosc2 Users's Manual, LSO--MAN--ESO--36100--0004

\bibitem[]{} Pozzetti, L., Madau, P., Zamorani, G., Ferguson, H.C., 
\& Bruzual, A.G. 1998, \mnras, 298, 1133

\bibitem[]{} Quataert, E., \& Narayan, R. 1999, \apj, 520, 298 

\bibitem[]{} Savage, B.D., \& Mathis, J.S. 1979, \araa, 17, 73 

\bibitem[]{} Schmidt M., et al. 1998, \aap, 329, 495 

\bibitem[]{} Snowden, S.L. 2002, in ``New Visions of the X-ray Universe
in the XMM--{\it Newton} and {\it Chandra} era'', in press

\bibitem[]{} Tozzi, P., et al. 2001, \apj, 562, 42

\bibitem[]{} Van der Bliek, N.S., Manfroid, J., \& Bouchet, P. 1996, \aaps, 119, 547

\bibitem[]{} Vignali, C., Comastri, A., Fiore, F., \& La Franca, F. 2001, \aap, 370, 900

\bibitem[]{} Vignati, P., et al. 1999, \aap, 349, L57 
\end{thebibliography}
\end{document}